\documentstyle[aps]{revtex}
\begin{document}
\input epsf.tex

\title{Single particle Green's function in the Calogero-Sutherland model
for rational couplings $\beta=p/q$.}
\author{D. Serban${}^{*}$, F. Lesage${}^{**}$, V. Pasquier.${}^*$}
\address{${}^{**}$Department of physics, University of Southern California,
Los-Angeles CA90089-0484.\\${}^*$ Service de Physique Th\'eorique,
CEN-Saclay, F91191 Gif-sur-Yvette, France}
\date{\today , USC-95-023}
\maketitle

\begin{abstract}
We derive an exact expression for the single particle Green
function in the Calogero-Sutherland model for all the
rational
values of the coupling $\beta$.
The calculation is based on Jack polynomial
techniques and the results are given in the
thermodynamical limit.
Two type of intermediate states contribute.
The first one consists of
a particle propagating out of the Fermi
sea and the second one consists of a
particle propagating in one direction, $q$ particles in the
opposite direction and $p$ holes.

\end{abstract}

\section{Introduction}

The Calogero-Sutherland model describes interacting
particles on a circle \cite{calogero,sutherland}.
Recent interest in the model arose from its relation
with the Haldane-Shastry chain ($\beta=2$)
and also with the random matrix
theory. Moreover, it has proven to be a good model to
study fractional statistics.

There has been several works devoted to derive the
correlation functions in the Calogero-Sutherland model
\cite{forr,forr1,nous,hz,ha,tata,min}.
In particular, the density-density correlation
functions were obtained.
The Green function where the annihilation operator acts
before the creation operator was also computed.
Recently, Zirnbauer and Haldane obtained the Green
function with the creation operator acting first
for the special values $\beta=1/2,1,2$
\cite{hazi}. In this paper we compute this Green
function in a simple way for all the rational values of $\beta$
using Jack polynomial techniques.
The method we use is closly related to our preceeding paper
\cite{nous}, but the computation is more involved.
There are also new physical phenomenas:
Contrary to what one might naively expect, the intermediate
states do not consist only of one particle which propagates
out of the Fermi sea. There is an additional contribution
where one particle propagates in one direction,
$q$ particles in the opposite direction and $p$ holes.

The model is defined on a
circle of length $L$ and the Hamiltonian given by~:
\begin{equation}
\hat{H}=
-\sum_{j=1}^N {1\over 2}{d^2 \over dx_j^2} + \beta (\beta-1)
{\pi^2\over L^2} \sum_{i<j} {1\over \sin^2(\pi (x_i-x_j)/L)}
\end{equation}
where we use units in which $\hbar=m=1$.
Its eigenfunctions are known and written~:
\begin{equation}
\Psi(x_i)=\Delta^\beta(x) \Phi(x_i)
\end{equation}
with~:
\begin{equation}
\Delta(x)=\prod_{i<j} \sin\left( {\pi (x_i-x_j) \over L}\right)
\end{equation}
and $\Phi(x_i)$ is a symmetric function of the variables
$x_i$. The functions $\Phi(x_i)$
are
symmetric polynomials in the variables $z_j=e^{i 2 \pi x_j/L}$
and they are labeled
by a partition $\lambda$. They are known as the
Jack polynomials $J_\lambda(z_i;\beta)$. The properties of
these polynomials were
recently
studied in the mathematical literature
\cite{macdo,stanley,kadell}.

The emergence of fractional statistics is seen by considering
the spectrum of $\hat H$. This spectrum is like a free spectrum
and is described by a set of quasi-momenta $k_i$.
The momentum and energy are given by additive laws~:
\begin{equation}
Q_\alpha=\frac{2\pi}{L} \sum_{i=1}^{N} k_i, \ \
E_\alpha=\frac{2\pi^2}{L^2} \sum_{i=1}^{N} k_i^2 ,
\end{equation}
and interaction between the particles is encoded in the fact
that the
quasi-momenta $k_i$ obey a generalized
exclusion principle~:
\begin{equation}
  k_{i+1}-k_i  = \beta+I_i.
\end{equation}
where $I_i$ are positive integers or zero.
The ground state is described by the configuration of $k_i$'s
the most densely packed around the origin, $I_i=0$.  For $N$
odd, the ground state momenta are~:
$$k_i^{(0)}=\beta \left( i-\frac{N+1}{2}\right).$$
By analogy with the fermion case ($\beta=1$) we
call this configuration the Fermi sea.
To describe the elementary excitations
it is convenient to multiply the quasi-momenta $k_i$
by $q$ so that they
differ by integer numbers.
Then, the Fermi sea can be described by a set of occupied
quasi-momenta separated by $p-1$ unoccupied ones.
In this occupation number representation, a particle corresponds to
one $1$ followed by $p-1$ zeroes and a hole corresponds
to a sequence of $q$ zeroes.
Note that the effect of removing $q$ particles from the Fermi sea
is to create $p$ holes (see figure 1).

 \vskip 0.5cm
\centerline{
\epsfxsize 12.0 truecm
\epsfbox{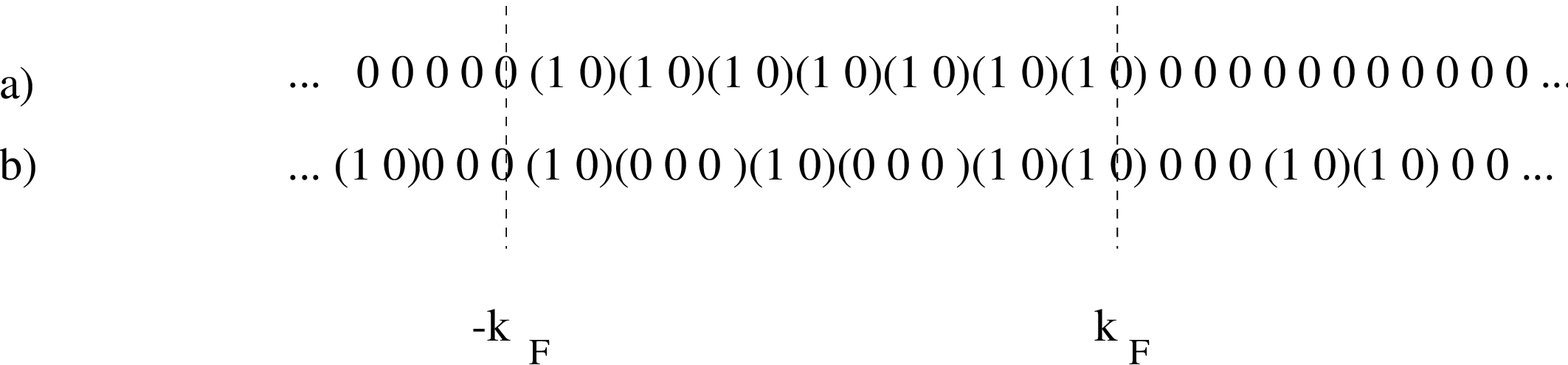}
}
\vskip 0.5cm
\centerline{{\bf Fig.1:}a) The ground state at $\beta=2/3$
b) an excitation
with $3$ particles and $2$ holes .}
\vskip 0.4cm

This description does not tell which type of excitations
are produced when one acts with a specific operator. This
information can only be obtained
by considering the properties of the
wave functions.
In the next part, we use these properties to obtain the
Green function form factor.

\section{Green function.}

The quantity of interest here is the Green function with the
creation operator acting before the annihilation operator.
The results in this section follow the paper  \cite{nous}
where the notations are defined.
It is written~:
\begin{eqnarray}
{}_N \! <0|\Psi(x,t) \Psi^\dagger(0,0)|0>_N&=\sum_{|\alpha>} { {}_N \!
<0|\Psi(x,t)|\alpha>_{N+1} {}_{N+1} \!
<\alpha | \Psi^\dagger(0,0) |0>_N \over {}_{N+1} \! <\alpha|\alpha>_{N+1}
{}_N \! <0|0>_N }
\nonumber \\
&=\sum_{|\alpha>} {\vert {}_{N} \! <0|\Psi(0,0)|\alpha>_{N+1}\vert^2
\over {}_{N+1} \! <\alpha|\alpha>_{N+1} {}_N \! <0|0>_N}
 e^{-iE_\alpha t
- i Q_\alpha x}
\end{eqnarray}
where $|\alpha>_{N+1}$ is a complete basis of states with
$N+1$ particles and $|0>_N$ has N particles.
This complete
set of states can be expressed in terms of Jack polynomials
in $N+1$ variables~:
\begin{equation}
|\alpha >_{N+1}=|n,\lambda>=
\prod_{i=1}^{N+1} z_i^n
J_\lambda(z_i;\beta)|0>_{N+1}
\end{equation}
with
\begin{equation}
|0>_{N+1}=\prod_{i=1}^{N+1} z_i^{-\beta (N+1)/2}
\prod_{i<j} (z_i-z_j)^\beta .
\end{equation}
and
\begin{equation}
|0>_{N}=\prod_{i=1}^N z_i^{-\beta (N-1)/2}
\prod_{i<j} (z_i-z_j)^\beta .
\end{equation}
$n$ is an
arbitrary integer and
the partition $\lambda$ is such that $\l(\lambda)<N+1$
in order to avoid double counting of states.
The quasi-momenta $k_i$ entering the expression
of the energy and momentum (4) of the state $|\alpha>$
are given in terms of
$\lambda$ by~:
\begin{equation}
k_i=\lambda_{N-i+1}+n+\beta \left(
i-\frac{N+3}{2}
\right).
\end{equation}

The vacuum state $|0>_{N+1}$ has been chosen in such a
way that the quasimomentum of the $N+1$ particle is at the left
of the Fermi sea of the $N$ particle vacuum. Another possible
choice would have been to put this particle to the right
by replacing the exponent of $z_i$ by $-\beta (N-1)/2$ in
the expression (8) of $|0>_{N+1}$. If $\beta$ is an integer,
these two states belong to the same Hilbert space and the two
definitions give the same form factor in the thermodynamical limit.
If $\beta$ is not integer, we shall see that the two possible choices
give different form factors in the thermodynamical limit.

The action of $\Psi(0,0)$ is to remove a particle at the positions
$x_i=0$, therefore setting the
variable $z_i$ to 1.  The matrix element we need to
compute is~:
\begin{equation}
M_\alpha={}_N \! <0|\prod_i z_i^{n-\beta} (1-z_i)^\beta
J_\lambda(1,z_j;\beta) |0>_N,
\end{equation}
for all $\lambda$.
It is possible to expand the Jack polynomial in $N+1$ variables
on polynomials in $N$ variables~:
\begin{equation}
J_{\lambda}(1,z_i;\beta)=\sum_\nu J_{\lambda/\nu}(1;\beta)
J_{\nu}(z_i;\beta)
\end{equation}
and here $J_{\lambda/\nu}(1;\beta)$ is different from zero only when
$\lambda$ covers $\nu$ and when the difference between $\lambda$ and
$\nu$ is at most a box per column.  When these conditions are
satisfied, the coefficient is given by\cite{stanley}~:
\begin{equation}
J_{\lambda/\nu}(1;\beta)=\prod_{s\in C_{\lambda/\nu}}
\left( \frac{\beta l(s)+a(s)+1}{\beta (l(s)+1)+a(s)}\right)_\lambda
\left( \frac{\beta (l(s)+1)+a(s)}{\beta l(s)+a(s)+1}\right)_\nu .
\end{equation}
The notations here are as follow: $s=(i,j)$ is a box on a
Young tableau identified by its coordinates $1\leq j \leq \lambda_i$.
Given a box, $s$, $a(s)$ is the number of boxes on its right and
$l(s)$ the number of boxes under it.  The notation $(...)_\lambda$
indicates that the quantities are evaluated with respect to the
partition $\lambda$ and $C_{\lambda/\nu}$ denotes the set of columns
of $\lambda$ which have the same length as the corresponding
column of $\nu$.

The product in (11) can also be expanded on Jack polynomials
with the relation \cite{kadell}~:
\begin{equation}
\prod_i (1-z_i)^\beta =\sum_\mu H^\beta(\beta;\mu)
J_\mu(z_i;\beta)
\end{equation}
with coefficient $H^\beta(\beta;\mu)$~:
\begin{equation}
H^\beta(\beta;\mu)=\prod_{s\in \mu} \frac{j-1-\beta i}{1+a(s)+\beta l(s)}
\end{equation}
with notations defined above.  The
results of \cite{kadell} are proven for $\beta$ integer
but it is simple to extend the proof to arbitrary rational
$\beta$ using Kaneko's generalized Selberg integrals \cite{kaneko}.
This coefficient
is zero whenever the partition $\mu$ has more than
$p$ legs or more than $(q-1)$ arms for $\beta=p/q$. This is
at the origin of the selection rule for the intermediate states.
Using these
expansions and the orthogonality relation for the Jack polynomials
\cite{macdo}~:
\begin{equation}
{}_N \! <0|J_\nu (z_i;\beta) J_\rho(\bar{z}_i;\beta)|0>_N=
\delta_{\nu,\rho} {\cal N}_\rho(\beta,N)
\end{equation}
we find the matrix element $M_\alpha$~:
\begin{equation}
M_\alpha=(-1)^{\beta N}
\sum_\nu H^\beta(\beta;\nu+n) J_{\lambda/\nu}(1;\beta)
{{\cal N}_\nu(\beta,N)\over \sqrt{{\cal N}_0(\beta;N)
{\cal N}_\lambda(\beta;N+1)}}
\end{equation}
where  $\nu+n$ is the partition $\nu$ with $n$
columns added (or subtracted when $n$ negative).
In this expression $\nu+n$ is a partition with
at most $p$ legs, therefore $n \leq p$.

The correlation function is now written~:
\begin{equation}
{}_N \! <0|\Psi(x,t)\Psi(0,0)^\dagger|0>_N=
\sum_{|\alpha >} |M_\alpha|^2  e^{iQ_\alpha x-i E_\alpha t}
{}.
\end{equation}
The evaluation of $M_\alpha$ in the thermodynamical
limit is our main result. All the quantities entering
(17) are known from the mathematical literature and
can be also found in \cite{nous}.

There
is a first contribution coming from $n=p$ and $n=-r$
which will be denoted $G^{(1)}$.  It this part, the only partitions
contributing correspond to Young Tableaux, $\lambda$, having
one ``arm" or one part, $\lambda=(\lambda_1)$, with $\lambda_1$ an arbitrary
positive integer or Young tableaux having $N$ parts of length $r$
denoted $\lambda=(r^N)$. Each type of tableaux
account for a particle propagating  on a different side of the
Fermi sea.
In the thermodynamical limit, we obtain:
\begin{equation}
G^{(1)}(x,t)=\rho_0
\beta \int_1^\infty dw \left( {w-1\over w+1}\right)^{\beta-1}
\ 2\cos(Q x) e^{-i E t}
\end{equation}
where we used the variable~:
\begin{equation}
{\lambda_1\over \beta N}={w-1\over 2}
\end{equation}
and
\begin{equation}
E={\rho_0^2 \beta^2 \pi^2\over 2}w^2 , \ \ Q=\rho_0 \beta \pi w,
\end{equation}
with $\rho_0=N/L$.
This is interpreted as the propagation of one particle over the
Fermi sea.  For $\beta=1$, the fermions, we find the correct
result.

The contributions for $n=1,...,(p-1)$ are present for finite
$N$ but in the thermodynamic limit they are suppressed by factors
of $N$.  The only remaining contribution is for $n \leq 0$.
It comes from partitions $\lambda$ which
are of the form~:
\begin{equation}
(r+\lambda_1,...,r+\lambda_q,
(r+p)^{\lambda_p'-q},(r+p-1)^{\lambda_{p-1}'
-\lambda_p'},
...,r^{N-\lambda_1'})
\end{equation}
\centerline{
\epsfxsize 8.0 truecm
\epsfbox{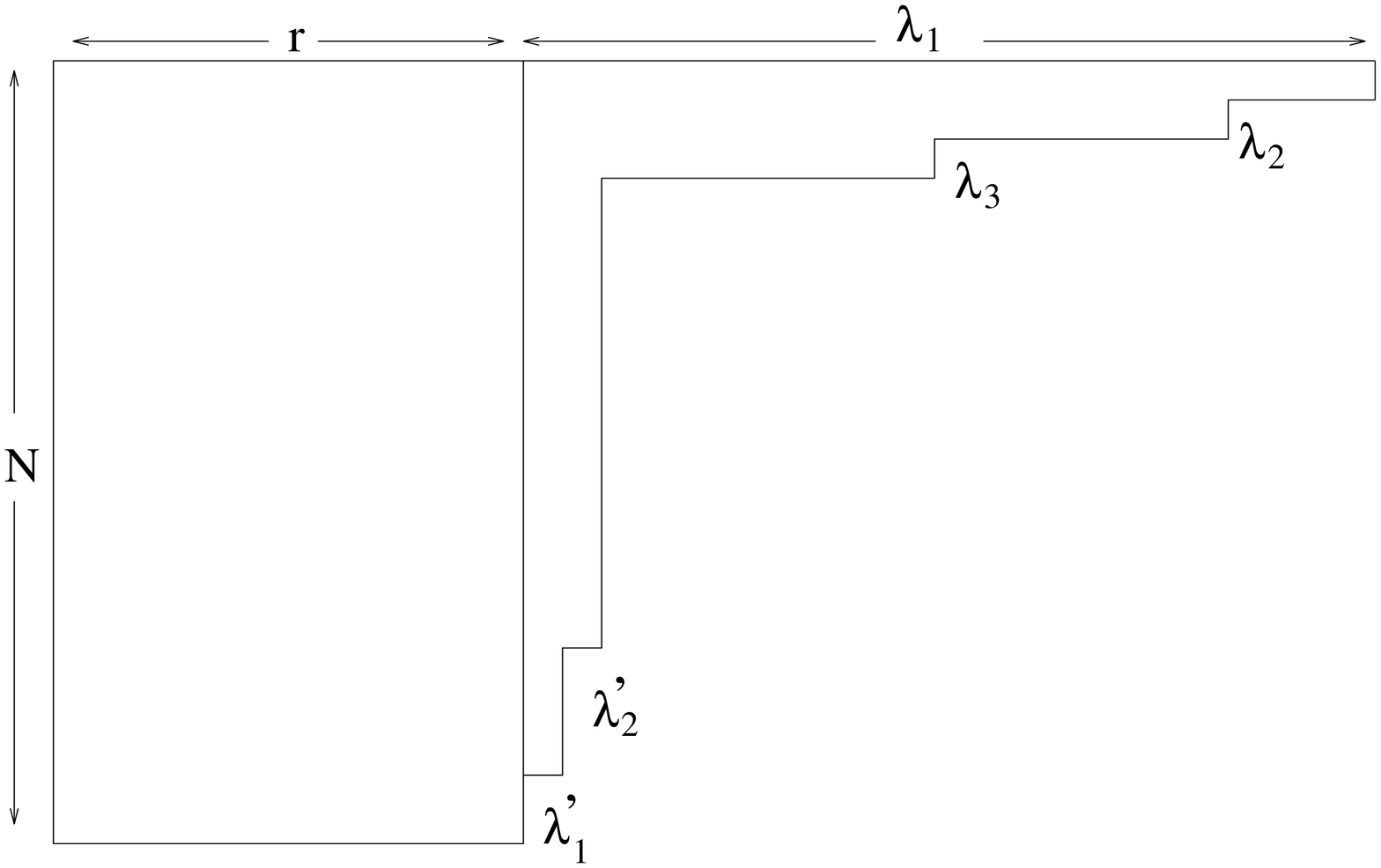}
}
\vskip 0.4cm
\centerline{{\bf Fig.2:} Young tableaux contributing to the correlation
for $\beta=2/3$.}
\vskip 0.4cm

Here the $\lambda_i$'s  $(i=1,..,q)$ and $r=-n$ are non negative
integers and correspond to $q$ particles propagating
in the forward direction and one particle
in the backward direction outside the Fermi sea .
The $\lambda_i'$ $(i=1,...,p)$
are bounded by
$N$ and are related to the momenta of the holes inside
the Fermi sea.

When we take the continuum limit with the following
variables~:
\begin{equation}
\frac{\lambda_i-(\beta-1)i}{\beta N}=\frac{w_i-1}{2},
\ \ \frac{\beta\lambda_i'-i}{\beta N}=
\frac{1-v_i}{2}, \ \ \frac{r}{\beta N}=-\frac{w_{0}+1}{2},
\ \ \frac{\nu_i-(\beta-1)i}{\beta N}=\frac{\xi_i-1}{2}
\end{equation}
we obtain the second part of the
correlation function .  The expression is given by~:
\begin{equation}
G^{(2)}(x,t)=C(\beta) \int_{1}^{\infty} \prod_{j=1}^q dw_j
\int_{-\infty}^{-1} dw_0
\int_{-1}^1 \prod_{i=1}^p dv_i \
F(v_i,w_j) \ e^{-i E_\alpha t - Q_\alpha x},
\end{equation}

The energy and momentum are~:
\begin{equation}
E_\alpha={\beta \pi^2 \rho_0^2\over 2} \left(
-\sum_{i=1}^p v_i^2+\beta \sum_{i=0}^qw_i^2\right)
 , \ \ Q_\alpha= \pi \rho_0 \left( \sum_{i=1}^p v_i -
\beta \sum_{i=0}^q w_i \right)
\end{equation}
and the constant~:
\begin{equation}
C(\beta)=\rho_0 \frac{\beta^{-pq+1} \Gamma(q/p)^p}{2p!q! \Gamma(p/q)^q}
\prod_{i=1}^p \Gamma(qi/p)^{-2} \prod_{i=1}^{q-1} \Gamma(-pi/q)^{-2} .
\end{equation}

The form factor $F(v_i,w_j)$ is equal to~;

\begin{equation}
F(v_i,w_j)={
\prod_{1\leq i<j \leq p} \vert v_i-v_j\vert^{2/\beta} \prod_{0 \leq i<j
\leq q}
\vert w_i-w_j\vert^{2\beta }
\over \prod_{i=1}^p (1-v_i^2)^{1-1/\beta}\
\prod_{j=0}^{q} (w_j^2-1)^{1-\beta}\prod_{i,j} (v_i-w_j)^2} {\cal K}^2,
\end{equation}

\begin{eqnarray}
&{\cal K}=  \frac{\prod_{j=1}^p \prod_{i=0}^q (w_i-v_j)^{\beta}}
{\prod_{k<j}(v_k-v_j)\prod_{0\leq i<j\leq q}(w_i-w_j)^{2\beta-1}}
\int_{w_q}^{w_{q-1}}  d\xi_{q-1} \cdots \int_{w_2}^{w_1}
d\xi_1 \prod_{i<j}^{q-1} (\xi_i-\xi_j) \prod_{i=1}^{q-1}
\prod_{j=0}^q (\xi_i-w_j)^{\beta-1} \nonumber \\
& \partial_{v_1}...\partial_{v_p} \left[ \prod_{k<j} (v_k-v_j)
\prod_{j=1}^p \left( \prod_{i=0}^q (w_i-v_j)^{1-\beta}
\prod_{i=1}^{q-1} (\xi_i-v_j)^{-1} \right) \right]
\end{eqnarray}

In the appendix we give the main lines of the derivation of ${\cal K}$
and we argue that this expression can be
simplified to a single contour integral~:
\begin{equation}
{\cal K}=(\beta-1)\frac{\Gamma^q(\beta)}{2\pi i}
\int_{{\cal C}_w}  dz \frac{\prod_{i=1}^p (v_i-z)}{
\prod_{j=0}^q (w_j-z)^\beta}
\end{equation}
where ${\cal C}_w$ is a contour surrounding the points $w_1,...,w_q$
as in the figure 3.

 \vskip 0.5cm
\centerline{
\epsfxsize 12.0 truecm
\epsfbox{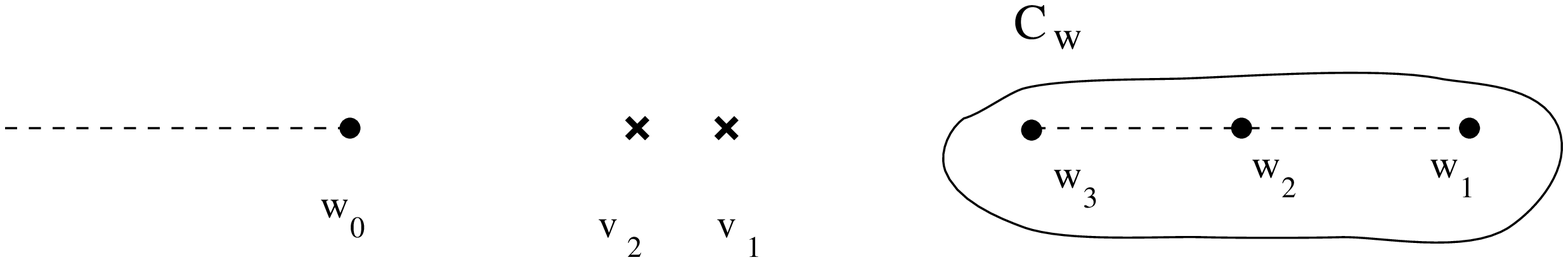}
}
\vskip 0.5cm
\centerline{{\bf Fig.3:} Contour of integration.
The branch cuts of the integrand are represented by dashed lines.}
\vskip 0.4cm

When $\beta=1/2$, this expression of ${\cal K}$ coincides with the result
obtained by Zirnbauer and Haldane \cite{hazi}.
For $\beta$ integer (or, equivalently, $q=1$)
this result is rigorous and proves the conjecture in \cite{hazi}.
The free fermion case $\beta=1$ is special in the sense
that the second contribution to the form factor vanishes.

We tested numerically the equality of the integrals (28)
and (29) for $\beta=1/2$ and $\beta=3/2$.

This completes the computation of the form factor of the
advanced Green function for all rational values of the
coupling constant $\beta$.

\section{Conclusions.}

We computed the advanced Green function in the Calogero-Sutherland
model using Jack polynomial techniques.  The computation done here
follows from the techniques already used in \cite{nous}.

The final expression for the form factor we obtain is the
same for $\beta$ integer or not. There is however a big difference
between the two cases. If $\beta$ is an integer, the form factor
is invarient under a parity transformation ( change of sign of
all the quasimomenta which define the intermediate state $\alpha$).
If $\beta$ is not an integer, one particle propagates in the forward
direction and q particle propagate in the backward direction
but there is no contribution coming from a particle propagating
in the backward direction and q particles in the forward direction.
The form factor is thus not invariant under a parity transformation.
The origin of this phenomena can be traced back to two possible
inequivalent $N+1$ particles vacua. Had we chosen the other possibility,
we would have obtained the form factor conjugated under parity.
The physical meaning of this phenomena is not clear to us.

The results presented here
simplify drastically in the continuum limit.
Finding a way to
do these computations directly in this limit would be of great
interest.  The complexity of the part described by ${\cal K}$ in our
results show that simple bosonic vertex operators don't seem to
reproduce the right answer.
Note however that the integrals over the variables $\xi$ in the first
expression of ${\cal K}$
bear striking similarities with the screening operators which
occur in the Coulomb Gaz representation of conformal field theories
\cite{dots}

\section{Acknowledgements.}

We wish to thank M.R. Zirnbauer for discussions about the form factors
and M. Bauer and S. Nonnenmacher for helping us with the complex
analysis.
F. Lesage is supported by a Canadian NSERC 67 Scholarship and a
Canadian Postdoctoral Fellowship.

\appendix

\section{Summation.}

Let us go back to the expression for the form factor~:
\begin{equation}
M_\alpha=(-1)^{\beta N}
\sum_\nu H^\beta(\beta;\nu+n) J_{\lambda/\nu}(1;\beta)
{{\cal N}_\nu(\beta,N)\over \sqrt{{\cal N}_0(\beta;N)
{\cal N}_\lambda(\beta;N+1)}}
\end{equation}
and let us look at this expression for a fixed partition
$\lambda$ and a negative integer $n$.  Then the coefficient
$H^\beta(\beta;\nu+n)$ forces the partition $\nu+n$ to
have $p$ legs and $(q-1)$ arms.  Since $n$ is negative,
the partition $\nu$ must have $r=-n$ columns and then
$p$ legs and $(q-1)$ arms.

The main difficulty of the computation is to make the
summation in (A1) on the partitions $\nu$ such that
$J_{\lambda/\nu}\neq 0$.
As explained in \cite{nous} , only
cases for which
$\lambda_i'-\nu_i'=0,1$ are allowed.  The case
considered in figure 2 corresponds to $\beta=2/3$.
Again, we characterize the partition $\lambda$ by
the numbers $r,\lambda_i,\lambda'_j,$
which define the quasi-particle momenta in the continuum limit.
In this case the partition $\lambda$ is of the form~:
$$
(r+\lambda_1, r+\lambda_2, r+\lambda_3, (r+2)^{\lambda_2'-3},
(r+1)^{\lambda_1'-\lambda_2'}, r^{N-\lambda_1'}),
$$
and $\nu$ can be read similarly from the figure~:
$$
(r+\nu_1, r+\nu_2,(r+2)^{\nu_2'-2},(r+1)^{\nu_1'-\nu_2'},r^{N-\nu_1'}).
$$
\centerline{
\epsfxsize 8.0 truecm
\epsfbox{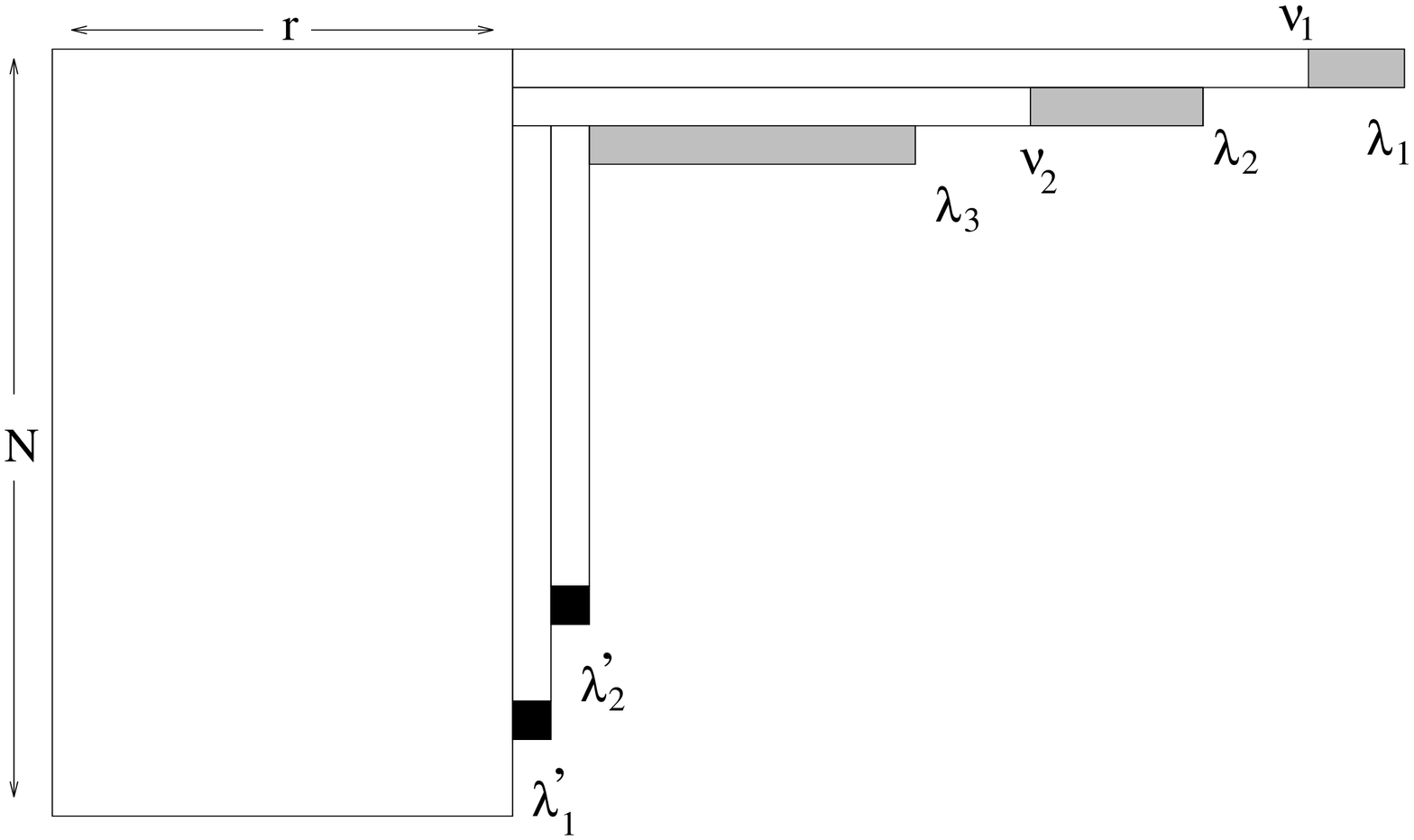}
}
\vskip 0.5cm
\centerline{{\bf Fig.4:} Young tableaux setting the notation.}
\vskip 0.4cm
When we keep $\nu_i$ fixed, we must sum over the possibilities
$\nu_i'=\lambda_i' , \lambda_i'-1$ corresponding to
the black boxes in the figure.
The restriction that $\lambda/\nu$ is a one dimensional
strip gives the following constraints on the $\nu_i$'s~:
\begin{equation}
\lambda_2 \leq \nu_1 \leq \lambda_1 , \ \
\lambda_3 \leq \nu_2 \leq \lambda_2 .
\end{equation}

If we use the variables appropriate for the continuum limit
\begin{equation}
\frac{\lambda_i-(\beta-1)i}{\beta N}=\frac{w_i-1}{2},
\ \ \frac{\beta\lambda_i'-i}{\beta N}=
\frac{1-v_i}{2}, \ \ \frac{r}{\beta N}=-\frac{w_{0}+1}{2},
\ \ \frac{\nu_i-(\beta-1)i}{\beta N}=\frac{\xi_i-1}{2},
\end{equation}
and we will use $a=\frac{2}{\beta N}$.
The sum over the $\nu_i$'s in the
intervals defined by (A2) become integrals over the continuum variables
$\xi_i$`s.  This correspond
to the shaded parts in the figure.

Let us concentrate on the sum over $\nu'_i$, factorizing out the part
which is independent of $\nu'_i$.
This sum results in the following expression~:
\begin{equation}
Q_N=\frac{1}{a^p} \sum_{I} (-1)^{| I |}
\prod_{i\in I}\left[\prod_{k=0}^q
\frac{w_k-v_i-(\beta-1) a}{w_k-v_i}
\prod_{k=1}^{q-1} \frac{\xi_k-v_i-a}{
\xi_k-v_i} \prod_{j\not\in I}
\frac{v_j-v_i+a}{v_j-v_i}\right].
\end{equation}
Here $I$ is the ensemble for which
$\lambda_i'=\nu_i'$ and we have to sum over all
possibilities $I\subset\{1,...,p\}$.
The symbol $|I|$ means the cardinal of the
ensemble.
Before taking the thermodynamical limit of this expression, we
note the following property~:
\begin{equation}
\lim_{a\to 0}\frac{1}{a^p}
\sum_{I} (-1)^{| I |}  \prod_{i\in I} \left[
\frac{f (v_i-a)}{f (v_i)}
\prod_{j\not\in I} \frac{v_j-v_i+a}{v_j-v_i}\right] =
\frac{\partial_{v_1}...\partial_{v_p}\prod_{i<j}(v_i-v_j)
\prod_{i=1}^pf(v_i) }{\prod_{i<j}(v_i-v_j)\prod_{i=1}^pf(v_i) }
\end{equation}
This expression,which comes from the definition of
the derivative, can be also written as~:
\begin{eqnarray}
\lim_{a_1,...,a_p\to0} \frac{1}{a_1...a_p} \sum_{I} (-1)^{| I |}
\prod_{i\in I} \frac{f(v_i-a_i)}{f(v_i)}
\prod_{i\in I,j\in I}\left(\frac{v_j-v_i-a_j+a_i}{v_j-v_i}\right)
\nonumber \\
\prod_{i\in I,j\not\in I}\left(\frac{v_j-v_i+a_i}{v_j-v_i}\right)
\prod_{i\not\in I,j\in I}\left(\frac{v_j-v_i-a_j}{v_j-v_i}\right)
\end{eqnarray}
and it is unchanged if a quantity of order one in $a_i$ is added.
Let $f(v)$ have the form $f(v)=f_1^{\alpha_1}(v)f_2^{\alpha_2}(v)$.
To first order in $a_i$~:
$$\frac{f(v_i-a_i)}{f(v_i)}\approx
\frac{f_1(v_i-\alpha_1a_i)f_2(v_i-\alpha_2a_i)}
{f_1(v_i)f_2(v_i)}=
1-a_i\left(\alpha_1\frac{f'_1(v_i)}{f_1(v_i)}+\alpha_2\frac{f'_2(v_i)}
{f_2(v_i)}\right)+{\cal O}(a_i^2).$$
Replacing this in (A5) and putting $a_1=...=a_p=a$ gives
the large $N$ limit of the sum in (A3), with $f_1=\prod_{i=1}^q
(w_i-v)$, $f_2=\prod_{i=1}^{q-1} (\xi_i-v)$ and
$\alpha_1=1-\beta$, $\alpha_2=-1$.

\begin{equation}
Q= \frac{
\partial_{v_1}...\partial_{v_p} \left[ \prod_{k<j} (v_k-v_j)
\prod_{j=1}^p \left( \prod_{i=0}^q (w_i-v_j)^{1-\beta}
\prod_{i=1}^{q-1} (\xi_i-v_j)^{-1} \right) \right]}
{\prod_{k<j}(v_k-v_j)\prod_{j=1}^p \left[\prod_{i=0}^q (w_i-v_j)^{1-\beta}
\prod_{i=1}^{q-1} (\xi_i-v_j)^{-1} \right] }
\end{equation}

It remains to do the integrals over the variables
$\xi_i$'s. When we multiply (A7) by the part depending
explicitly on $\nu_i$
and factor out the part which is independent
of $\nu_i$  we obtain the following multiple integral~:
\begin{eqnarray}
&I(\beta)=  \frac{\prod_{j=1}^p \prod_{i=0}^q (w_i-v_j)^{\beta-1}}
{\prod_{k<j}(v_k-v_j)}
\int_{w_q}^{w_{q-1}}  d\xi_{q-1} \cdots \int_{w_2}^{w_1}
d\xi_1 \prod_{i<j}^{q-1} (\xi_i-\xi_j) \prod_{i=1}^{q-1}
\prod_{j=0}^q (\xi_i-w_j)^{\beta-1} \nonumber \\
& \partial_1...\partial_p \left[ \prod_{k<j} (v_k-v_j)
\prod_{j=1}^p \left( \prod_{i=0}^q (w_i-v_j)^{1-\beta}
\prod_{i=1}^{q-1} (\xi_i-v_j)^{-1} \right) \right]
\end{eqnarray}
Doing explicitly these
integrals is not an easy task. However, a careful analysis
of this expression suggests it can take the simpler form
of a single contour integral.
In the following we present the arguments leading to
this conjecture.

The integral $I$ can be regarded as a complex function of
the variables $w_0,...,w_q$ and $v_1,...,v_p$.
We try to obtain the limit of $I$ when
the points $w_1,...,w_q$ collapse onto
the point $w$ and $v_1,...,v_p\to v$. Due to the
antisymmetry of the integrand as a function of $\xi_1,...,\xi_{q-1}$,
changing counterclockwise $w_{i-1}$ and $w_i$, $(i=2,...,q)$
just multiplies the integral by a phase ${\rm e}^{i\pi(2\beta-1)}$.
We expect then branch points at $w_i=w_j$.
The singular part of the integral responsible of these
branch points can be calculated:

\begin{eqnarray}
I_0&=&
\int_{w_q}^{w_{q-1}}  d\xi_{q-1} \cdots \int_{w_2}^{w_1}
d\xi_1 \prod_{i<j}^{q-1} (\xi_i-\xi_j) \prod_{i=1}^{q-1}
\prod_{j=1}^q (\xi_i-w_j)^{\beta-1}
\nonumber \\
&=&\frac{\Gamma^q(\beta)}{\Gamma(q\beta)}
\prod_{1\leq i<j\leq q} (w_i-w_j)^{2\beta-1}R(w_1,...,w_q)
\end{eqnarray}
The function $R(w_1,...,w_q)$ is symmetric, rational,
homogeneous of order zero and it has no singularities, so it is
constant. Evaluating the integral in the
'nested' limit $w_1\to (w_2\to...\to (w_q\to w)...)$
gives $R(w_1,...,w_q)=1$.

 Further simplification arises
when we specialize the variables $v_i=v$. This
operation is delicate when done on an expression like (A7),
so it is easier to restart from the discrete version (A4)
and put $v_i=v+ai$. The sum reduces over the ensembles $I=\{1,...s\},
\ s\leq p$~:
\begin{eqnarray}
Q_N&=& \frac{1}{a^p}
\sum_{s=0}^p (-1)^s {p \choose s}
\prod_{i=1}^s
\left[\prod_{k=0}^q
\frac{w_k-v-ai-(\beta-1) a}{w_k-v-ai}
\prod_{k=1}^{q-1} \frac{\xi_k-v-ai-a}{
\xi_k-v-ai}\right] \nonumber \\
&=&\frac{1}{G(v)} \sum_{s=0}^p (-1)^s {p \choose s}
G(v+as)
\end{eqnarray}
where $$G(v)= \prod_{k=0}^q\left(\frac{w_k-v}
{a}-\beta+1\right)_{\beta-1}
\prod_{k=1}^{q-1} \left( \frac{\xi_k-v}{a}-1\right)$$
Here the notation $(x)_n=\Gamma(x+n)/\Gamma(x)$ is used.
In the last term of the equality (A10) one recognizes the discrete
$p^{\rm th}$ derivative of $G(v)$. Using Stirling's formula and the
homogeneity of $G$ we get in the
large $N$ limit~:
\begin{equation}
\frac{(-1)^{p}}{\prod_{i=0}^q (w_i-v)^{\beta-1}
\prod_{i=1}^{q-1} (\xi_i-v)}
\frac{d^p}{dv^p} \left[ \prod_{i=0}^q (w_i-v)^{\beta-1}
\prod_{i=1}^{q-1} (\xi_i-v) \right] .
\end{equation}
Note that the powers $\beta-1$ and $1$ have opposite
sign in
(A7) and (A11).
Now, we can take the limit of $I$ when the $w_i$'s collapse
to $w$ and all the $v_i$'s are set equal to $v$.
Without the constant and the factor $\prod_{i<j}(w_i-w_j)^{2\beta-1}$,
the result is~:
\begin{equation}
(w-w_0)^{(\beta-1)(q-1)} (w-v)^{1-qp} \frac{d^p}{dv^p} \left[
(w-v)^{p-1} (w_0-v)^{p/q-1}\right]
\end{equation}
and this can be rewritten in the following form~:
\begin{equation}
(\beta-1) (w_0-w)^{2p-q} (w-v)^{-pq} (w_0-v)^{-p}
\frac{d^{p-1}}{dw^{p-1}} \left[ \frac{(v-w)^p}{(w_0-w)^\beta}\right].
\end{equation}
The equality of the two formulas can be checked by expanding
the products under the derivatives and using the properties of the
binomial coefficients.
Then we write the derivative as a contour integral~:
\begin{equation}
\frac{d^{p-1}}{dw^{p-1}} \left[ \frac{(v-w)^p}{(w_0-w)^\beta}\right]=
\frac{(p-1)!}{2\pi i} \int_{{\cal C}_w}
dz \frac{(v-z)^p }{(z-w)^p (w_0-z)^\beta} .
\end{equation}
The conjecture is that we can undo the specialization
of the variables $w_j$ and $v_i$ to obtain

\begin{equation}
I=(\beta-1)\frac{\Gamma^q(\beta)}{2\pi i}
\prod_{0\leq i<j}^q (w_i-w_j)^{2\beta-1} \prod_{i=1}^p \prod_{j=0}^q
(w_j-v_i)^{-1} \int_{{\cal C}_w}
dz \frac{\prod_{i=1}^p (v_i-z)}{\prod_{i=1}^q (z-w_i)^\beta (w_0-z)^\beta}
\end{equation}
where ${\cal C}_w$ is a contour in the complex plane encircling
$w_1,...,w_q$.
This function has the good analyticity properties
we can expect from (A8).

For $\beta$ integer (or $q=1$) this result is rigorous.
In this case, there is no variable of integration $\xi$
and $Q$ has simple poles at $w_i=v_j$, the residu being a
polynomial of order $2q-1=1$ in each $v_i$. This justify the
passage from (A13) to (A15).
$\beta=1/2$ is another case to test our conjecture. The contour
of integration ${\cal C}_w$ can be deformed to obtain ${\cal C}_{w_0}$.
As $\beta<1$ we have to subtract the divergent contribution
of the contour at infinity, we thus obtain~:
\begin{equation}
I=\frac{1}{4}\prod_{i=0}^2 \frac{1}{w_i-v} \int_{-\infty}^{w_0}
\frac{dy}{\sqrt {w_0-y}}\left(\frac{v-y}{\sqrt {(w_1-y)(w_2-y)}}-1\right)
\end{equation}
which is the result Zirnbauer and Haldane obtained
using a supersymmetric calculation \cite{hazi}.

\end{document}